\documentclass[fleqn,10pt]{wlscirep}
\usepackage{fontspec}

\usepackage{unicode-math}
\usepackage{setspace}
\usepackage{color}
\usepackage{multirow}
\usepackage{caption}
\usepackage{fontsize}
\title{A Unified Graph Neural Network Framework for Non-Equilibrium Carrier and Lattice Dynamics Driven by Electric Fields}

\author[1]{Jia-Wen Li}
\author[2,3,4,*]{Sheng Meng}
\author[1,+]{Xinghua Shi}
\author[1,$\ddagger$]{Jin Zhang}
\author[5]{Wei-Hai Fang}

\affil[1]{Laboratory of Theoretical and Computational Nanoscience, National Center for Nanoscience and Technology, Chinese Academy of Sciences, Beijing 100190, China}
\affil[2]{Beijing National Laboratory for Condensed Matter Physics, and Institute of Physics, Chinese Academy of Sciences, Beijing 100190, China}
\affil[3]{School of Physical Sciences, University of Chinese Academy of Sciences, Beijing 100049, China}
\affil[4]{Songshan Lake Materials Laboratory, Dongguan, Guangdong 523808, China}
\affil[5]{College of Chemistry, Key Laboratory of Theoretical and Computational Photochemistry of Ministry of Education, Beijing Normal University, Beijing 100875, China}

\affil[*]{smeng@iphy.ac.cn}
\affil[+]{shixh@nanoctr.cn}
\affil[$\ddagger$]{jinzhang@nanoctr.cn}

\begin{abstract}\normalsize

Finite-temperature simulations of electric-field-driven dynamics need a unified description of interatomic interactions, local electronic states, and configuration-dependent electric responses.
First-principles simulations remain scale-limited, whereas conventional machine-learning potentials lack electric-field effects.
Recent machine-learning frameworks have incorporated electric-field response or atom-resolved electronic-state information, but rarely both within a single framework.
Here, we develop an electric-field-response graph neural network (EFR-GNN) that predicts energies, forces, Born effective charge tensors, atom-resolved charges and magnetic moments, and supports long-time field-driven molecular dynamics with atom-resolved tracking of localized electronic states.
In hole-doped MgO, static fields rectify thermally activated hole-polaron hopping through a forward--backward asymmetry quantified by a nearest-neighbor model. 
In GaAs, resonant terahertz excitation generates a coherent Γ-point transverse-optical phonon with dephasing consistent with experiment, while opposite helicities reverse its rotation. 
In superionic $\alpha$-AgI, it reproduces temperature-dependent Ag$^+$ mobility and collective field-driven ionic drift.
Together, EFR-GNN offers an approach to finite-temperature simulations of field-driven atomic and localized-carrier dynamics.
\end{abstract}
\begin{document}

\flushbottom
\maketitle
\thispagestyle{empty}

\newpage



Electric fields offer a general and effective route to driving material dynamics, whose evolution is jointly governed by thermal fluctuations, interatomic interactions, local electronic states, and configuration-dependent electric responses \cite{Lu2017a,Basini2024,Wu2022a}.
Their simulation requires an accurate potential-energy surface together with a consistent description of local electronic states and electric-field-induced forces.
First-principles calculations provide descriptions of material--field interactions, yet their computational cost largely limits accessible system sizes and timescales \cite{Unke2021,Noe2020,Onida2002,Casida2012,Curchod2018,Ruggenthaler2018}.
Machine-learning interatomic potentials (MLIPs) substantially extend the length and time scales accessible to finite-temperature atomistic simulations by reproducing density functional theory (DFT) energies and forces at much lower cost, while conventional MLIPs remain largely limited to field-free dynamics \cite{Batatia2025,Chen2026a,Chen2026,Zhang2018,Batzner2022,Batatia2022}.

Building on this foundation, recent machine-learning frameworks for atomistic dynamics have expanded along two capability dimensions: electric-field response and atom-resolved electronic-state information, as shown in Figure \ref{Fig1}(b).
Electric-field-responsive MLIPs incorporate configuration-dependent electric responses and their coupling to external fields into the learned representation, thereby extending machine-learning molecular dynamics (MLMD) to field-driven dynamics and enabling the prediction of related electric-response quantities.
Existing formulations encode electric coupling either directly in a field-dependent generalized potential, whose derivatives determine forces and Born effective charges (BECs) \cite{Falletta2025,Zhang2023b}, or through predicted BECs coupled to a field-free machine-learning force field \cite{Joll2024,Yu2025,Lu2026,Kutana2025,Zhong2025,Fan2026}. 
Within the latter class, BEC tensors are predicted explicitly \cite{Joll2024,Yu2025,Lu2026,Kutana2025}, or inferred from polaron or latent charges \cite{Zhong2025,Fan2026}.
These electric-field-responsive formulations provide complementary routes for incorporating configuration-dependent electric responses into MLMD.

In parallel, charge-aware models directly learn atom-resolved electronic quantities from first-principles labels, such as local magnetic moments, oxidation states, and charge populations, to resolve local valence states and atom-resolved charge distributions \cite{Deng2023,Eckhoff2020,Birschitzky2025,Ahart2026,Ko2021}.
Notably, although many models also employ atom-resolved charge variables, these quantities are often introduced as latent variables for energy corrections or for reproducing polarization and BECs, rather than being directly supervised by first-principles atomic charges \cite{Fan2026,Unke2019,Zhong2025}.
Despite their dependence on the charge-partitioning scheme, DFT-supervised atom-resolved charges provide useful descriptors of local valence, charge redistribution, and carrier localization.
Polaron-specific implementations further learn atom-resolved charges and local magnetic moments to identify the instantaneous carrier location, follow its transfer between atomic sites, and analyze thermally activated hopping over extended MLMD trajectories \cite{Birschitzky2025,Ahart2026}.
These charge-aware frameworks support long-time simulations and atom-resolved analysis of localized-carrier dynamics, but their applications remain largely field-free. 
Field-responsive atomistic frameworks and charge-aware models have therefore extended MLMD along complementary but largely separate capability dimensions, with few frameworks incorporating both.



To connect these two capability dimensions, we develop an electric-field-response graph neural network (EFR-GNN) with first-principles-supervised atom-resolved charge and local-magnetic-moment prediction, as illustrated in Fig.~\ref{Fig1} (a).
The model adopts a modular, parameter-separated architecture with branches for the field-free potential-energy surface, atom-resolved charges and magnetic moments, and explicitly predicted BEC tensors. 
This design allows each physical target to be optimized independently and activated according to the requirements of a given application.
The field-free branch determines the conservative atomic forces, the response branch predicts BEC tensors that convert the prescribed electric field into configuration-dependent field-induced forces, and the electronic-state branch predicts atom-resolved Bader charges and local magnetic moments that identify carrier localization and transfer.
EFR-GNN therefore combines configuration-dependent field-induced forces with optional atom-resolved electronic-state resolution within one MLMD framework.


Three systems were selected to represent complementary classes of field-driven nonequilibrium dynamics: hole-doped MgO for polaron transport, GaAs for terahertz-driven coherent lattice dynamics, and superionic $\alpha$-AgI for ionic migration, as shown in Fig.~\ref{Fig1}(c--e).
First, in hole-doped MgO, we reproduce thermally activated hole-polaron transport, which changes from stochastic motion to sustained drift under a static electric field.
This field-induced transport originates from a directional imbalance between forward and backward hopping events and is captured by a nearest-neighbor statistical model.
Time-dependent electric fields provide additional control through their tunable frequency, envelope, phase, polarization, and helicity, yet remain comparatively less explored in MLMD. 
This regime is examined in our second application to GaAs, where EFR-GNN reproduces the finite-temperature softening of the Γ-point transverse-optical phonon, selectively excites this mode with a linearly polarized terahertz pulse, and yields coherent-phonon dephasing times consistent with experiment. 
Circularly polarized pulses with opposite helicities further generate coherent rotating phonon states with opposite senses of rotation.
Third, in superionic $\alpha$-AgI, we capture the experimental temperature trend and activation energy of Ag$^+$ mobility, and directly simulate ionic drift under static electric fields.
By connecting configuration-dependent electric responses with atom-resolved charge information, EFR-GNN supports finite-temperature atomistic simulations of field-driven polaron transport, coherent lattice control, and ionic motion, opening access to a broader class of field-driven nonequilibrium processes.

\section*{Results}

\subsection*{Machine-learning framework}

\begin{figure}[hbpt]
	\centering
	\includegraphics[width=0.99\columnwidth]{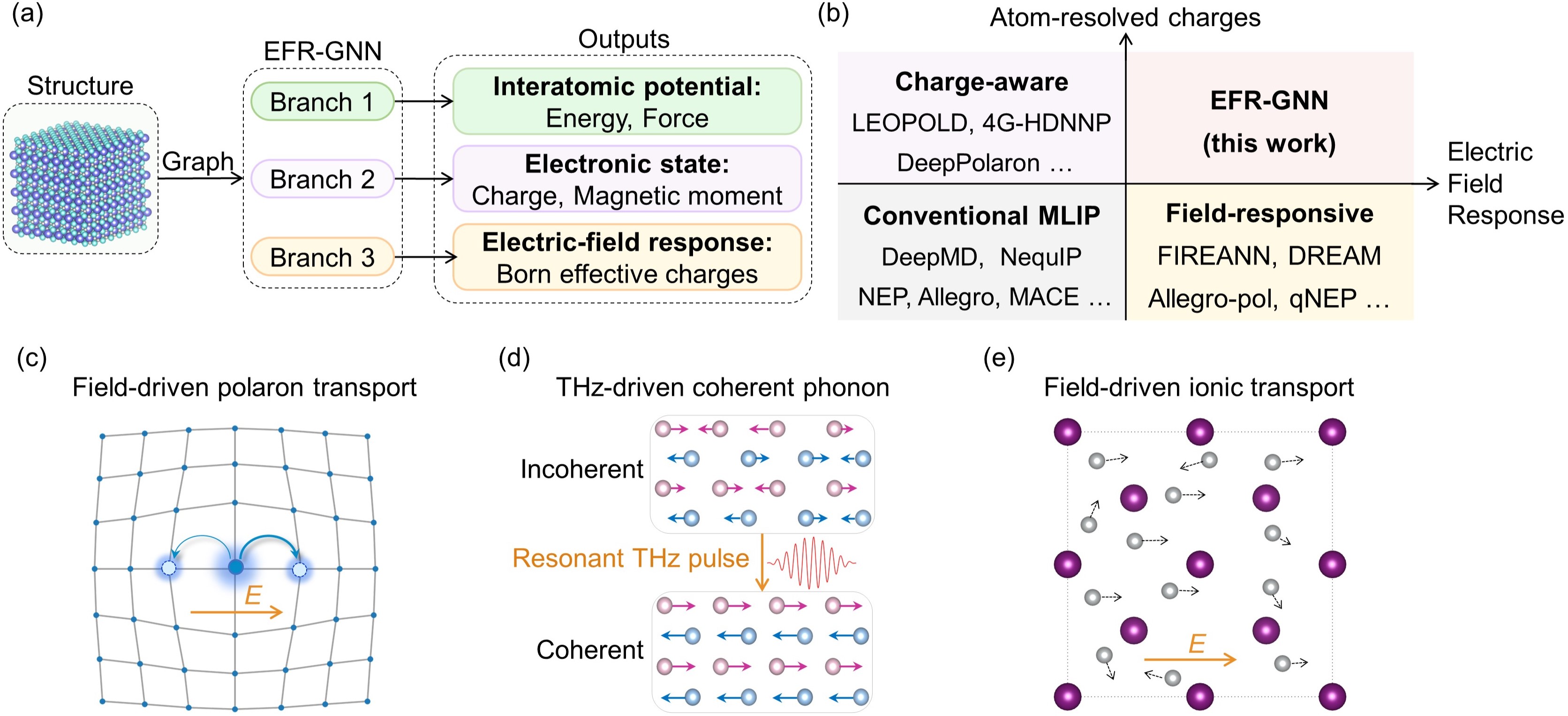}\\
	\caption{
		\textbf{Architecture and representative applications of the electric-field-response graph neural network (EFR-GNN).}
		(a) Atomic configurations are mapped onto graphs and processed by three parameter-separated equivariant branches that predict the field-free energy and forces, atom-resolved charges and magnetic moments, and Born effective charge tensors.
 		Branches 2 and 3 are optionally activated depending on the problem of interest.
 		(b) Capability map of representative machine-learning atomistic models.
 		Methods are categorized according to two demonstrated capabilities: electric-field response and first-principles-supervised prediction of atom-resolved charges.
 		Based on conventional machine-learning interatomic potentials (MLIPs), electric-field response extends machine-learning molecular dynamics (MLMD) to field-driven dynamics, whereas atom-resolved charge prediction supports atom-resolved tracking of charge distributions. 
 		Representative methods are listed \cite{Batatia2025,Chen2026a,Chen2026,Zhang2018,Batzner2022,Falletta2025,Zhang2023b,Joll2024,Yu2025,Lu2026,Kutana2025,Zhong2025,Fan2026,Deng2023,Eckhoff2020,Birschitzky2025,Ahart2026,Batatia2022}.
		(c--e) Schematic illustrations of electric-field-driven polaron transport, resonant terahertz (THz) excitation of a coherent phonon, and field-driven ionic transport, respectively.
	}
	\label{Fig1}
\end{figure}

The modular architecture of EFR-GNN is illustrated in Fig.~\ref{Fig1}. 
Each structure is represented as a graph, with atoms as nodes and neighboring atomic pairs within a cutoff radius as edges. 
The resulting graph is processed by three parameter-separated branches constructed from equivariant graph message-passing layers, which preserve the geometric transformation properties required for learning scalar, vector, and tensorial quantities \cite{Geiger2022,Wang2024a,Yang2024b,Zhong2026,Musaelian2023,Gong2023,Hsu2026,Li2026}.
Each branch further incorporates configurable node and edge updates, bond-type-aware attention, and trainable distance-dependent interaction attenuation.
Starting from the shared graph representation, the three branches are assigned distinct physical tasks.


Branch 1 predicts the field-free potential energy, whose negative gradient with respect to the atomic positions gives the corresponding field-free force on each atom.
Branch 2 predicts Bader charges and atomic magnetic moments, which serve as complementary descriptors for identifying localized electronic states and tracking their transfer between atomic sites.
Branch 3 predicts the atom-resolved Born effective charge tensors, whose product with the applied electric-field vector gives the field-induced force on each atom.
For charge-neutral systems, the predicted BEC tensors are corrected to satisfy the acoustic sum rule, thereby preserving translational invariance of the polarization response and eliminating the net force under a uniform electric field. 
The total atomic force is the sum of the field-free and field-induced forces.
Branches 2 and 3 are optional and can be activated according to the specific problem. 
The parameter separation further permits the three physical targets to be optimized independently, preventing gradient conflicts between the corresponding tasks.
Branches 1 and 3 determine the field-driven atomic trajectory, whereas Branch 2 predicts the atomic charges and magnetic moments along that trajectory.
A detailed description of the EFR-GNN architecture is provided in Supplementary Note 1 \cite{SI}.

\subsection*{Electric-field-driven small polaron dynamics in MgO}

Polaron transport involves the migration of a spatially localized electronic state and therefore requires atom-resolved predictions of charges and magnetic moments. 
Rocksalt MgO provides a prototypical platform, where an introduced hole localizes on an O anion and forms a small polaron \cite{Nabi2025}. 
Previous first-principles calculations placed the nearest-neighbor O--O hopping barrier near 0.1 eV \cite{Falletta2022,Smith2017}. 
Recent MLMD further identified thermally activated hopping along symmetry-equivalent [110] directions as the dominant equilibrium transport mechanism \cite{Birschitzky2025}. 


To describe finite-temperature atomic motion while tracking polaron localization, EFR-GNN was trained using 500 K first-principles molecular-dynamics trajectories in the canonical (NVT) ensemble that extensively sampled polaron hopping in MgO.
The model yields mean absolute errors (MAEs) of 0.039~meV atom$^{-1}$ and 13.7~meV\,\AA$^{-1}$ for energy and forces, respectively, as detailed in Supplementary Note~2 \cite{SI}.
Fig.~\ref{Fig2}(b--d) further shows the predicted Bader charges of the O atoms, local magnetic moments, and BEC components, with corresponding MAEs of 0.0041~e, $6.5\times10^{-4}$~$\mu_{\mathrm B}$, and 0.027~e. 
Polaron localization is identified by an electron depletion relative to the nonpolaronic O sites together with a finite local magnetic moment on the hosting O atom. 
The remaining O atoms retain nearly their reference Bader electron charges and negligible magnetic moments, as illustrated in Fig.~\ref{Fig2}(b,c). 
Accordingly, the instantaneous polaron position was tracked using the O atom with the maximum local magnetic moment. 
In addition, EFR-GNN reproduces the hopping energy profile while resolving the site-to-site transfer of the localized hole.
For a nearest-neighbor O--O transition, the predicted energy profile reproduces the DFT reference and gives a barrier of approximately 85 meV, as shown in Fig.~\ref{Fig2}(e), comparable in magnitude to the previously calculated value of about 110~meV \cite{Falletta2022,Smith2017}. 
As the polaron crosses the barrier, the local magnetic moment is rapidly transferred from the initial O site to its neighbor, as shown in Fig.~\ref{Fig2}(f), revealing the microscopic transfer of the localized hole.
Additional model parameters and performance evaluations are provided in Supplementary Note 2 \cite{SI}.

Zero-field polaron transport was subsequently evaluated using 500-ps MLMD trajectories in the microcanonical (NVE) ensemble at 200, 300, 400, and 500~K within Atomic Simulation Environment (ASE) \cite{Martyna1992NHC,HjorthLarsen2017}.
A $2\times2\times2$ supercell with 64 atoms was considered.
At each temperature, three independent trajectories were initialized with the polaron localized at different O sites, and the mobility was extracted from the polaron mean-squared displacements (MSDs) through the Einstein relation. 
The three trajectories yield closely comparable mobilities; accordingly, Fig.~\ref{Fig2}(g) shows their mean values.
Detailed MLMD settings and analyses of trajectory stability, polaron diffusion, and field-driven transport are provided in Supplementary Notes 3 and 4 \cite{SI}.
The resulting mobilities fall within the previously reported small-polaron mobility range of $10^{-3}\sim1$ cm$^2$V$^{-1}$s$^{-1}$ \cite{Smith2017,Birschitzky2025,Rettie2015,FUJISHIMA2008,Ahart2026}.
Nearly all hopping events occur between nearest-neighbor O sites. 
Consistent with this microscopic mechanism, an Arrhenius fit to the temperature-dependent mobility yielded an activation energy of 85~meV, matching the calculated nearest-neighbor hopping barrier in Fig.~\ref{Fig2}(e).

\begin{figure*}[hbpt]
	\centering
	\includegraphics[width=0.85\columnwidth]{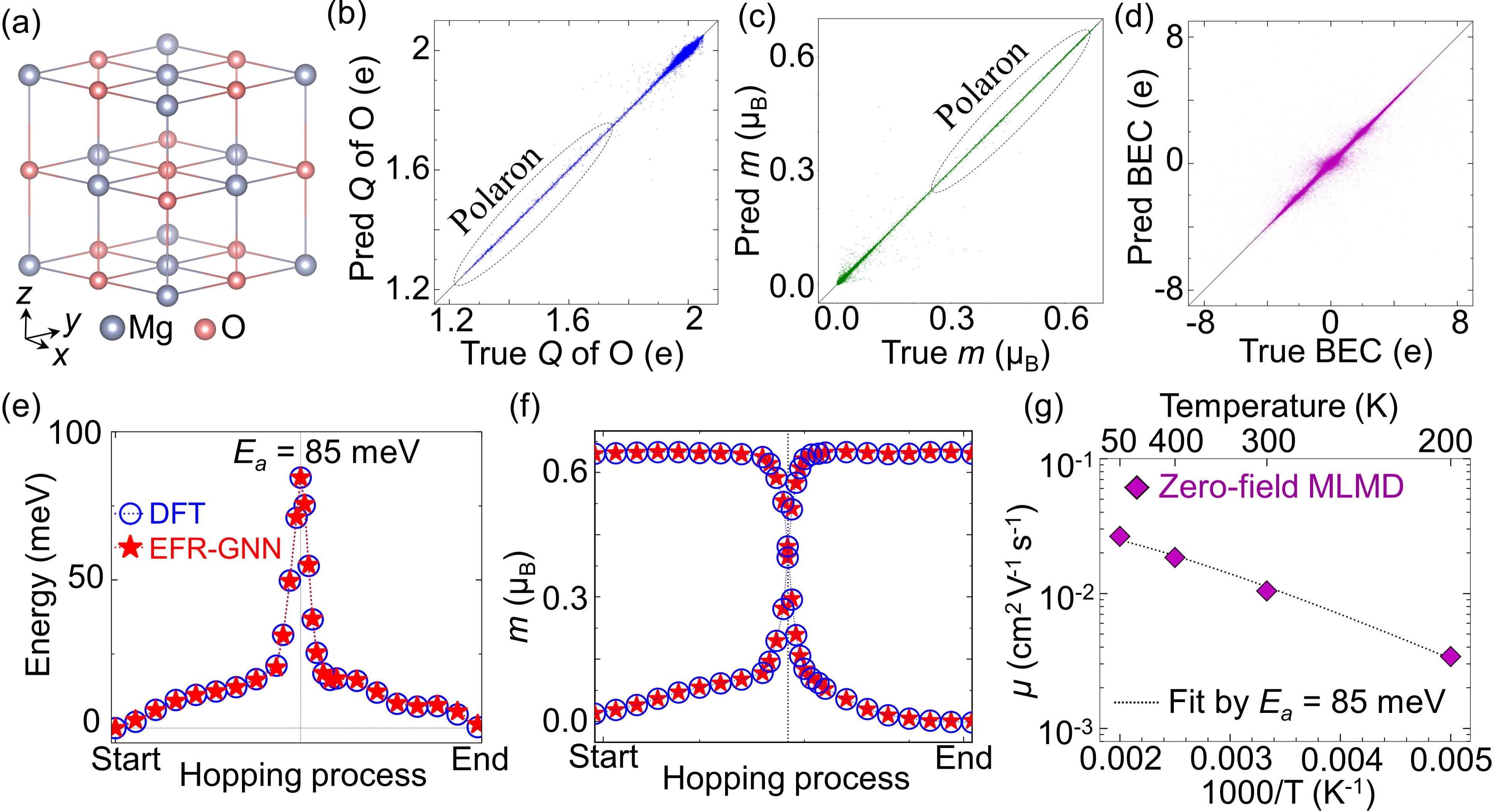}\\
	\caption{
		\textbf{EFR-GNN reproduces thermally activated hole-polaron hopping in MgO.}
		(a) Crystal structure of MgO.
		(b--d) Predicted versus density functional theory (DFT) reference values for the O-site Bader charges, local magnetic moments, and Born effective charge (BEC) components, respectively.
		The corresponding mean absolute errors (MAEs) are 0.0041~$e$, $6.5\times10^{-4}~\mu_{\mathrm B}$, and 0.027~$e$, respectively.
		A localized hole polaron has a reduced O electron charge and a finite local magnetic moment.
		(e) DFT and EFR-GNN energy profiles along a nearest-neighbor polaron-hopping pathway, yielding a barrier of approximately 85~meV.
		(f) Local magnetic moments along the hopping pathway, showing transfer of the polaron between neighboring O sites.
		(g) Temperature-dependent zero-field hole-polaron mobility from zero-field machine-learning molecular dynamics (MLMD).
		The Arrhenius fit gives $E_a$ = 85 meV, close to the static barrier in (e).
	}
	\label{Fig2}
\end{figure*}

\begin{figure}[hbpt]
	\centering
	\includegraphics[width=0.61\columnwidth]{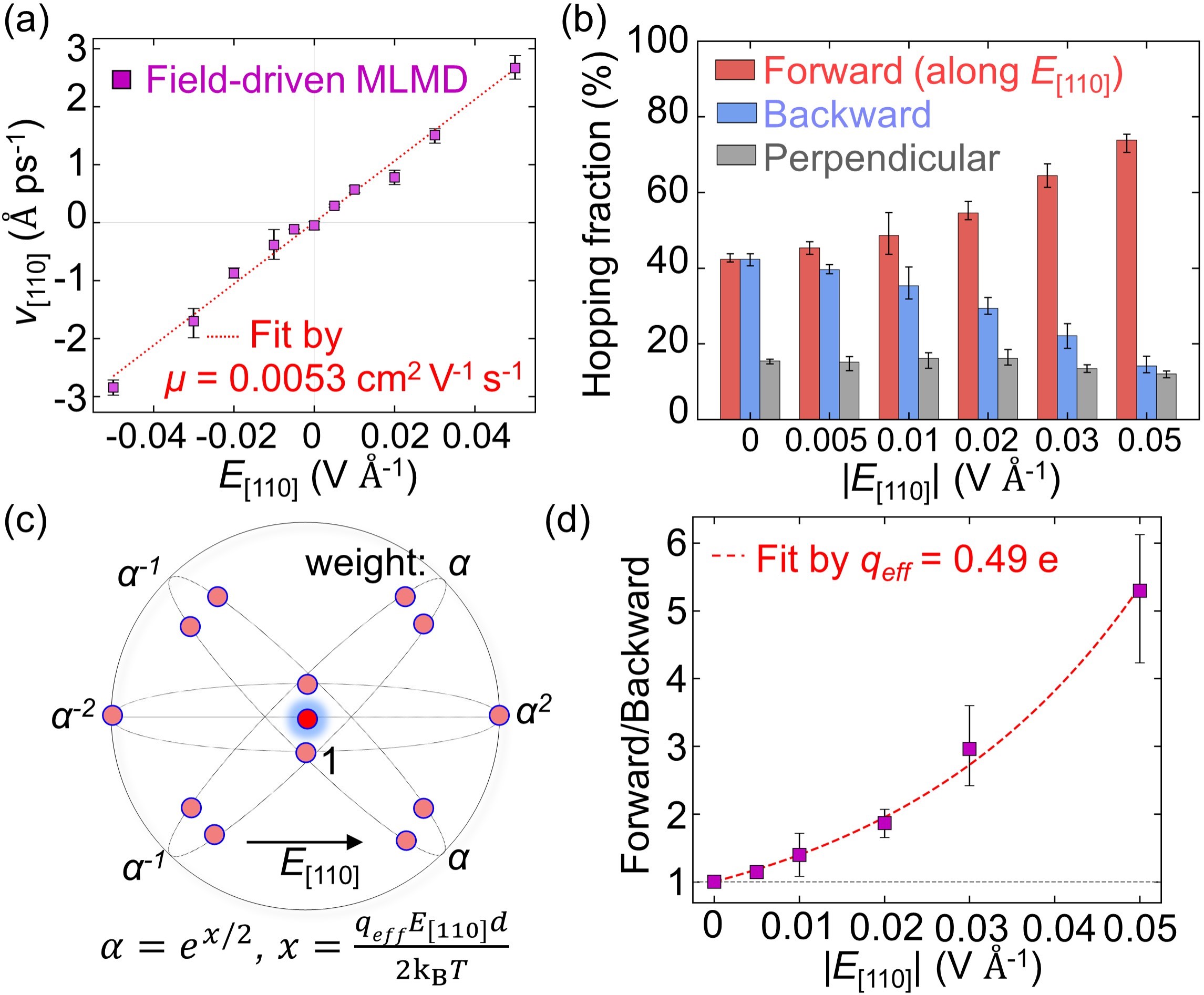}\\
	\caption{
		\textbf{Electric-field-driven hole-polaron transport in MgO.}
		(a) Hole-polaron drift velocity along $[110]$ as a function of the applied electric field.
		The linear fit gives a mobility of 0.0053 cm$^2$ V$^{-1}$ s$^{-1}$, consistent in magnitude with the zero-field value in Fig. \ref{Fig2}(g).
		(b) Fractions of forward, backward, and perpendicular hopping events; increasing the field enhances forward hopping and suppresses backward hopping.
		(c) Statistical model for the 12 nearest-neighbor hopping sites of O atom under a field along $[110]$, where the corresponding hopping weights are indicated.
		(d) Forward-to-backward hopping ratio fitted by the model in (c), giving an effective driving charge of 0.49 e.
		Error bars indicate the full range from the minimum to the maximum across all independent trajectories.
	}
	\label{Fig3}
\end{figure}

Static electric fields were next applied along the nearest-neighbor O--O hopping direction, [110], to directly probe field-driven hole-polaron transport.
Field strengths of $E_{[110]}=\pm0.005$, $\pm0.010$, $\pm0.020$, $\pm0.030$, and $\pm0.050$~V\,\AA$^{-1}$ were considered.
For each nonzero field strength, two independent 200-ps Langevin NVT MLMD trajectories at 300 K were initialized with the polaron localized at different O sites \cite{VandenEijnden2006Langevin}.
The zero-field result was obtained from three 500-ps NVE trajectories used for the equilibrium transport analysis in Fig.~\ref{Fig2}(g).
The finite-field mean drift velocity reverses with the field polarity and exhibits an approximately linear dependence on $E_{\rm[110]}$, as shown in Fig.~\ref{Fig3}(a). 
A linear fit yields a drift mobility of 0.0053~cm$^2$\,V$^{-1}$\,s$^{-1}$, in good agreement with the zero-field mobility independently obtained from the polaron MSD in Fig.~\ref{Fig2}(g).

To resolve the microscopic origin of the long-range drift, nearest-neighbor O--O hopping events under an electric field were classified by the projection of their displacements onto the field direction.
At zero field, the measured forward and backward hopping fractions are nearly identical, and the combined forward-to-backward-to-perpendicular ratio $5.00:5.01:1.82$ is consistent with the geometrical expectation of $5:5:2$.
With increasing field, forward hopping is enhanced and backward hopping is suppressed, as shown in Fig.~\ref{Fig3}(b).
Thus, electric-field-driven polaron transport in MgO originates from a field-induced forward bias that breaks the symmetry between forward and backward nearest-neighbor hopping.

The field-induced directional bias can be rationalized using the 12 nearest-neighbor O--O hopping directions of the face-centered-cubic oxygen sublattice illustrated in Fig.~\ref{Fig3}(c). 
For a hop along $\vec{d}_i$ under $\vec{E}_{\rm[110]}$, the electrostatic energy difference between the two O sites is
$\Delta U_i=-q_{\mathrm{eff}}\vec{E}_{\rm[110]}\cdot\vec{d}_i=-q_{\mathrm{eff}}E_{\rm[110]}d p_i$,
where $p_i=\vec{E}_{\rm[110]}/E_{\rm[110]}\cdot\vec{d}_i/d_i$ takes the values $1$, $1/2$, $0$, $-1/2$, and $-1$, with degeneracies of 1, 4, 2, 4, and 1, respectively. 
Then the field-dependent activation barrier is
$E_{\mathrm a,i}=E_{\mathrm a}^{0}-q_{\mathrm{eff}}E_{\rm[110]}d p_i/2$,
and the corresponding hopping weight is proportional to
$exp[q_{\mathrm{eff}}E_{\rm[110]}d p_i/(2k_{\mathrm B}T)]$. 
Defining $\alpha=exp[q_{\mathrm{eff}}E_{\rm[110]}d/(4k_{\mathrm B}T)]$, the directional weights are $\alpha^2$, $\alpha$, $1$, $\alpha^{-1}$, and $\alpha^{-2}$. 
After accounting for their degeneracies, the total forward, backward, and perpendicular weights are
$W_{\mathrm F}:W_{\mathrm B}:W_{\perp}=(\alpha^2+4\alpha):(\alpha^{-2}+4\alpha^{-1}):2$.
Here, $d=2.95$~\AA~is the nearest-neighbor O--O distance. 
At zero field, $\alpha=1$, giving the geometrical ratio
$W_{\mathrm F}:W_{\mathrm B}:W_{\perp}=5:5:2$.
As shown in Fig.~\ref{Fig3}(d), fitting the finite-field ratios yields $q_{\mathrm{eff}}=0.49$~$e$, comparable to the Bader electron depletion of approximately 0.6 e on the polaronic O site relative to nonpolaronic O sites.
The difference in $q_{\mathrm{eff}}$ arises because it also incorporates lattice relaxation, correlated hopping, recrossing, and variations in the hopping prefactor.
The simulations reveal that the electric field converts thermally activated hopping into sustained directional transport by breaking the forward--backward symmetry of nearest-neighbor hopping and establishing a net forward bias.

\subsection*{Terahertz excitation and helicity control of coherent transverse-optical phonons in GaAs}

\begin{figure*}[hbpt]
	\centering
	\includegraphics[width=0.91\columnwidth]{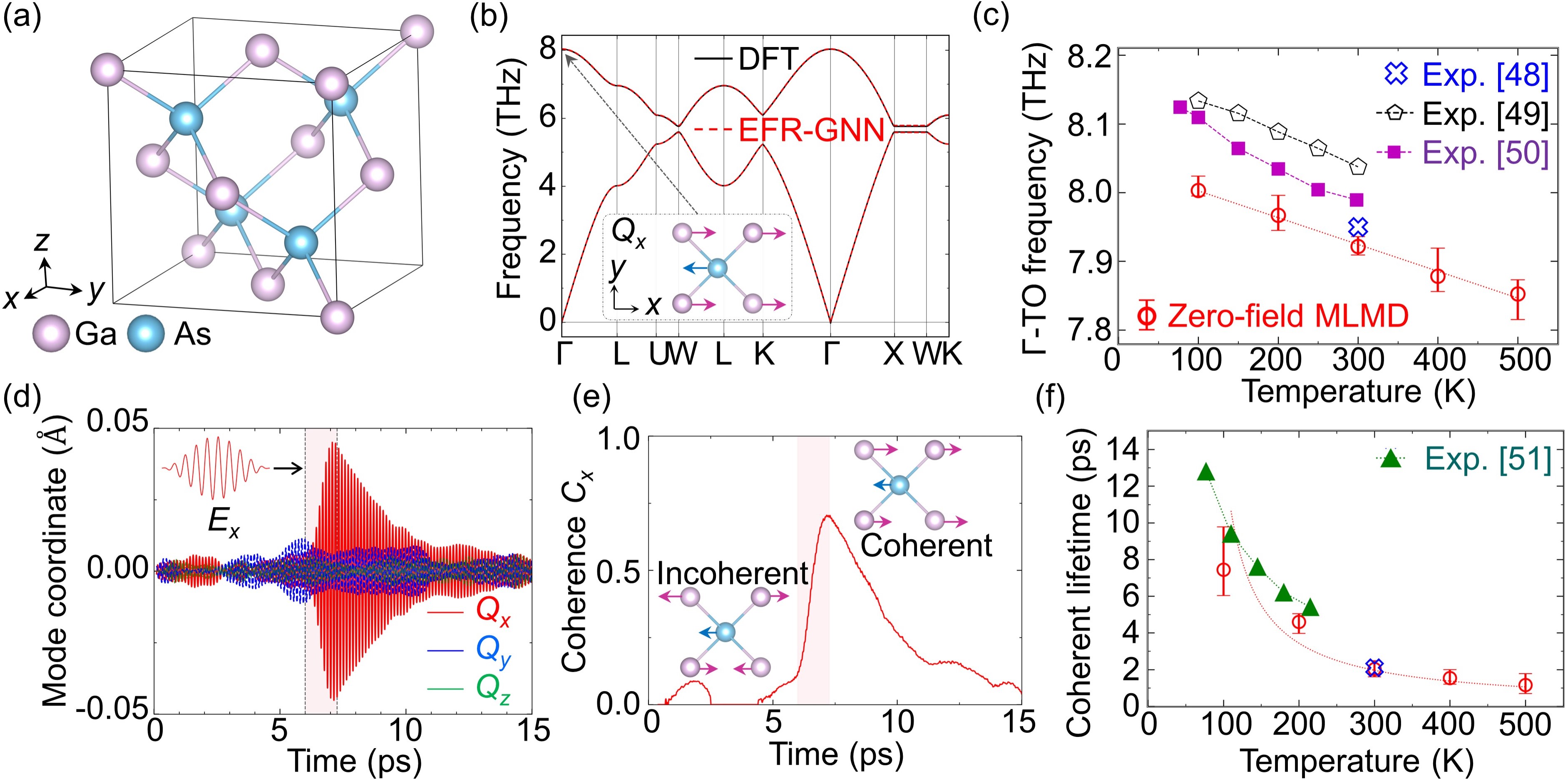}\\
	\caption{
		\textbf{Terahertz-driven excitation and dephasing of a coherent Γ-point transverse-optical phonon in GaAs.}
		(a) Crystal structure of zinc-blende GaAs.
		(b) Phonon dispersions calculated using DFT and EFR-GNN.
		The inset illustrates the $x$-polarized $\Gamma$-TO mode, $Q_x$, characterized by opposite displacements of the Ga and As sublattices.
		(c) Temperature-dependent Γ-TO frequency obtained from zero-field MLMD, compared with experiments \cite{Fu2016,Irmer1996,Komandin2024}.
		(d) Time evolution of the Cartesian mode coordinates $Q_x$, $Q_y$, and $Q_z$ under an $x$-polarized terahertz (THz) pulse at 300 K.
		The shaded region marks the pulse, which selectively excites $Q_x$ and is followed by post-pulse oscillation and decay.
		(e) Spatial coherence $C_x(t)$, showing the transition from incoherent thermal motion to a coherent state and its subsequent decay.
		(f) Temperature-dependent coherent-phonon dephasing time, which decreases with increasing temperature, and agrees well with experiments \cite{Fu2016,Vallee1991}.
	}
	\label{Fig4}
\end{figure*}

Time-dependent terahertz (THz) electric fields were next applied to resonantly excite polar phonons and control ultrafast lattice dynamics.
GaAs was selected as a prototypical III--V semiconductor with well-characterized polar optical phonons, and its zinc-blende structure is illustrated in Fig.~\ref{Fig4}(a).
The experimental cubic lattice constant of 5.65~\AA\ was adopted \cite{Blakemore1982}.
As illustrated in Fig.~\ref{Fig4}(b), GaAs hosts a threefold-degenerate transverse-optical (TO) phonon near 8~THz at the Brillouin-zone center (Γ), with components polarized along the $x$, $y$, and $z$ directions. 
Each component involves opposite motions of the Ga and As sublattices, with the $x$-polarized mode illustrated in the inset of Fig.~\ref{Fig4}(b). 
Its nonzero mode effective charge enables direct resonant coupling to THz electric fields, and coherent excitation of this phonon has been demonstrated experimentally \cite{Fu2016}.

The trained EFR-GNN model yields low MAEs of 0.04~meV atom$^{-1}$ for energy, 12.2~meV\,\AA$^{-1}$ for atomic forces, 0.0003 $e$ for Bader charge, and 0.073~$e$ for the BEC components.
The predicted phonon dispersion agrees with the DFT reference and reproduces the threefold-degenerate infrared-active Γ-TO mode near 8~THz, as shown in Fig.~\ref{Fig4}(b).
Detailed model parameters and evaluations are shown in Supplementary Note 5 \cite{SI}.

Beyond the harmonic dispersion, the finite-temperature anharmonic renormalization of the Γ-TO mode was examined under zero-field MLMD in a Nos\'e--Hoover-chain NVT ensemble \cite{Martyna1992NHC}, where a 512-atom supercell was used.
At each temperature, the renormalized TO frequency was assigned to the dominant peak of the Hann-windowed Fourier spectrum of the mode velocity.
The extracted frequency decreases monotonically from approximately 8.00~THz at 100~K to 7.79~THz at 500~K, as shown in Fig.~\ref{Fig4}(c). 
Both the absolute frequency scale and the temperature-induced redshift agree well with experiment \cite{Irmer1996,Fu2016,Komandin2024}, indicating that the model captures the anharmonic renormalization of the polar TO phonon.

We next examined the control of the Γ-TO phonon using a linearly polarized THz pulse. 
The electric field was prescribed as $E_x(t)=E_0s(t)\sin[\omega(t-t_0)]$, with $E_0=0.01$~V/\AA, $\omega=2\pi f$, $t_0=6$ ps and $f=7.9$~THz to match the finite-temperature $\Gamma$-TO frequency. 
The pulse envelope was defined as $s(t)=\sin^2[\pi(t-t_0)/T_{\mathrm p}]$ for $t_0\leq t\leq t_0+T_{\mathrm p}$ and zero otherwise, where $T_{\mathrm p}=10/f$ corresponds to 10 field cycles.
The THz-driven MLMD simulations were performed at 300~K in the NVT ensemble using a Nos\'e--Hoover-chain thermostat \cite{Martyna1992NHC,HjorthLarsen2017}. 
For each primitive-cell Ga--As basis pair $l$, $u_{s,l,\alpha}$ denotes the displacement of sublattice $s$ along the Cartesian direction $\alpha$. 
The local optical displacement was defined as $q_{l,\alpha}=u_{\mathrm{Ga},l,\alpha}-u_{\mathrm{As},l,\alpha}$, and the corresponding macroscopic coordinate as $Q_{\alpha}=N^{-1}\sum_lq_{l,\alpha}$, where $N$ is the number of Ga--As pairs.
As shown in Fig.~\ref{Fig4}(d), the pulse selectively excites $Q_x$, whereas $Q_y$ and $Q_z$ remain close to their thermal backgrounds, showing polarization-selective excitation of the Γ-TO mode.
In contrast, pulses at half and double frequencies (3.95~THz and 15.8~THz) produce negligible responses, supporting that the induced $Q_x$ oscillation arises from frequency-selective resonant excitation.

\begin{figure*}[hbpt]
	\centering
	\includegraphics[width=0.99\columnwidth]{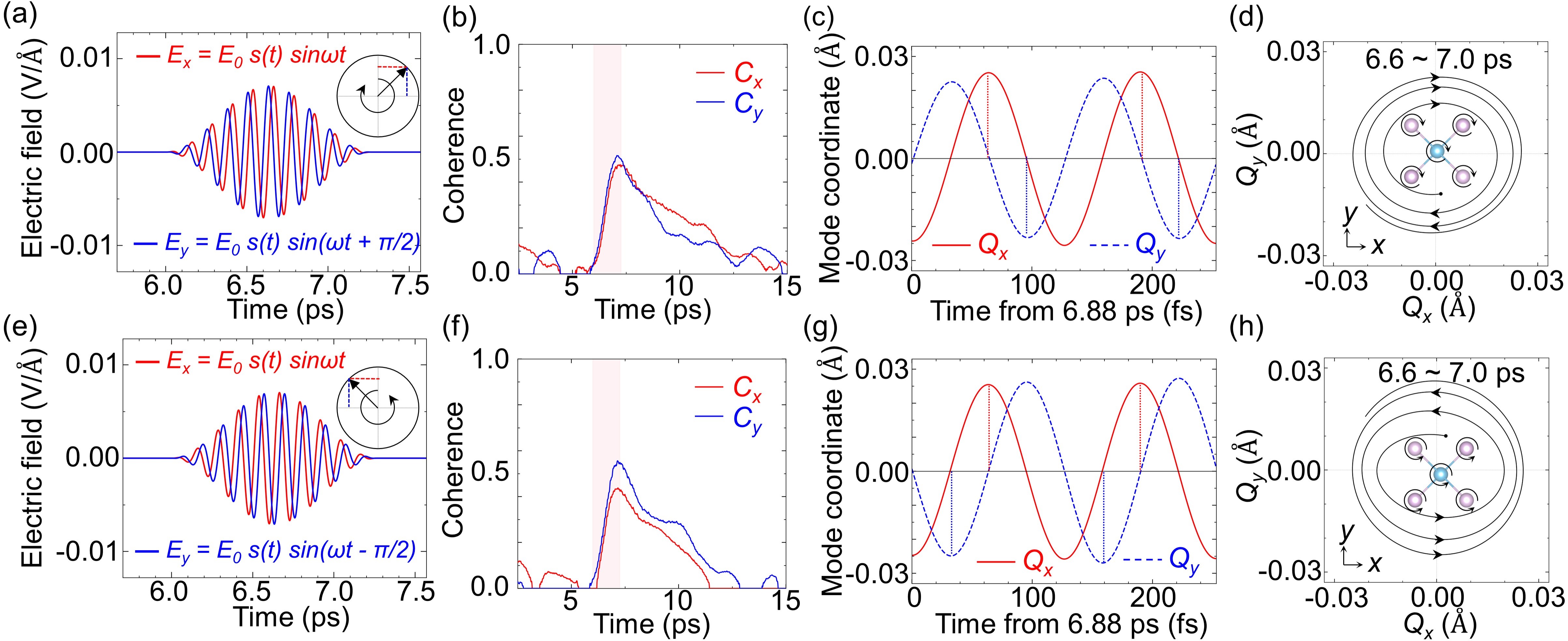}\\
	\caption{
		\textbf{Helicity-controlled reversal of coherent Γ-TO phonon rotation in GaAs.}
		(a,e) Orthogonal electric-field components of circularly polarized THz pulses with opposite helicities.
		(b,f) Corresponding spatial coherences $C_x$ and $C_y$ of the two degenerate Cartesian Γ-TO modes.
		The shaded regions mark the pulses; both coherences increase during excitation and decay afterward.
		(c,g) Comparable amplitudes and relative phases of approximately $\pm\pi/2$ produce coherent rotational lattice motion.
		(d,h) Viewed along the $+z$ direction, the resulting $Q_x$--$Q_y$ trajectories rotate in opposite directions, showing helicity-controlled reversal of the coherent lattice rotation.
	}
	\label{Fig5}
\end{figure*}

The driven $\Gamma$-TO phonon was characterized by its spatial coherence and dephasing time.
For each local Ga--As pair, the optical displacement was demodulated at the $\Gamma$-TO frequency within a centered time window to extract its instantaneous amplitude and phase.
The spatial coherence was then evaluated from the phase alignment of these local oscillations across the supercell, with the finite-size random-phase contribution removed.
As shown in Fig.~\ref{Fig4}(e), the corrected coherence $C_x(t)$ increases markedly during the THz pulse, indicating that the local Ga--As vibrations become synchronized to form a spatially coherent Γ-TO phonon.
The lifetime of this coherent state was quantified by fitting the post-pulse decay of the collective phonon amplitude to an exponential function.
As shown in Fig.~\ref{Fig4}(f), the resulting dephasing time decreases systematically with increasing temperature. 
At room temperature, the predicted value of $\sim$2.0~ps agrees quantitatively with the experimentally measured TO dephasing time of $\sim$2.1~ps \cite{Fu2016}, while the low-temperature values are comparable in magnitude to those reported for the longitudinal-optical mode \cite{Komandin2024}. 
These results show that EFR-GNN captures both the THz-driven formation of spatially coherent phonons and their temperature-dependent dephasing.
Detailed MLMD results and analyses are shown in Supplementary Notes 6 and 7 \cite{SI}.

Beyond the selective excitation of the single $Q_x$ component by a linearly polarized THz pulse, the degeneracy of the Γ-TO mode enables two orthogonal components to be coherently driven simultaneously by a circularly polarized THz field.
Ten-cycle circularly polarized pulses at 7.9 THz were constructed from two equal-amplitude orthogonal components,
$E_x(t)=E_0s(t)\sin[\omega(t-t_0)]/\sqrt{2}$ and
$E_y^{\pm}(t)=E_0s(t)\sin[\omega(t-t_0)\pm\pi/2]/\sqrt{2}$,
with $s(t)=\sin^2[\pi(t-t_0)/T_{\mathrm p}]$, $T_{\mathrm p}=10/f$, $t_0=6$ ps and $E_0=0.01$ V/\AA.
As illustrated in Fig.~\ref{Fig5}(a,e), the $+\pi/2$ and $-\pi/2$ phase shifts generate clockwise and counterclockwise rotating THz fields, respectively.

As a result, the simultaneous rise of $C_x$ and $C_y$ in Fig.~\ref{Fig5}(b,f) shows that $Q_x$ and $Q_y$ become coherent concurrently. 
Within the coherent window from 6.88 to 7.13 ps, $Q_x$ and $Q_y$ have comparable amplitudes and are offset by approximately one quarter of an oscillation period, as shown in Fig.~\ref{Fig5}(c,g).
The clockwise rotating field produces a $Q_y$-leading-$Q_x$ response, whereas the counterclockwise rotating field produces the opposite phase ordering. 
This phase-order reversal produces $Q_x$--$Q_y$ trajectories with opposite senses of rotation over 6.6--7.0~ps in Fig.~\ref{Fig5}(d,h), revealing a deterministic correspondence between the THz helicity and the rotational sense of the coherent lattice motion and providing a microscopic target for polarization-resolved electro-optic or magneto-optical measurements.
These results show direct THz control of polar lattice motion in GaAs, from resonant coherent-phonon excitation to the helicity-selective generation of coherent rotating phonon states.
Such helicity-controlled coherent lattice motion connects to broader efforts to manipulate phonon angular momentum and phonon-mediated magnetic responses \cite{Luo2023,Davies2024,Zhang2014,Nova2016,Afanasiev2021,Juraschek2025,Minakova2026}.


\subsection*{Electric-field-driven ion transport in superionic $\alpha$-AgI}

\begin{figure*}[hbpt]
	\centering
	\includegraphics[width=0.99\columnwidth]{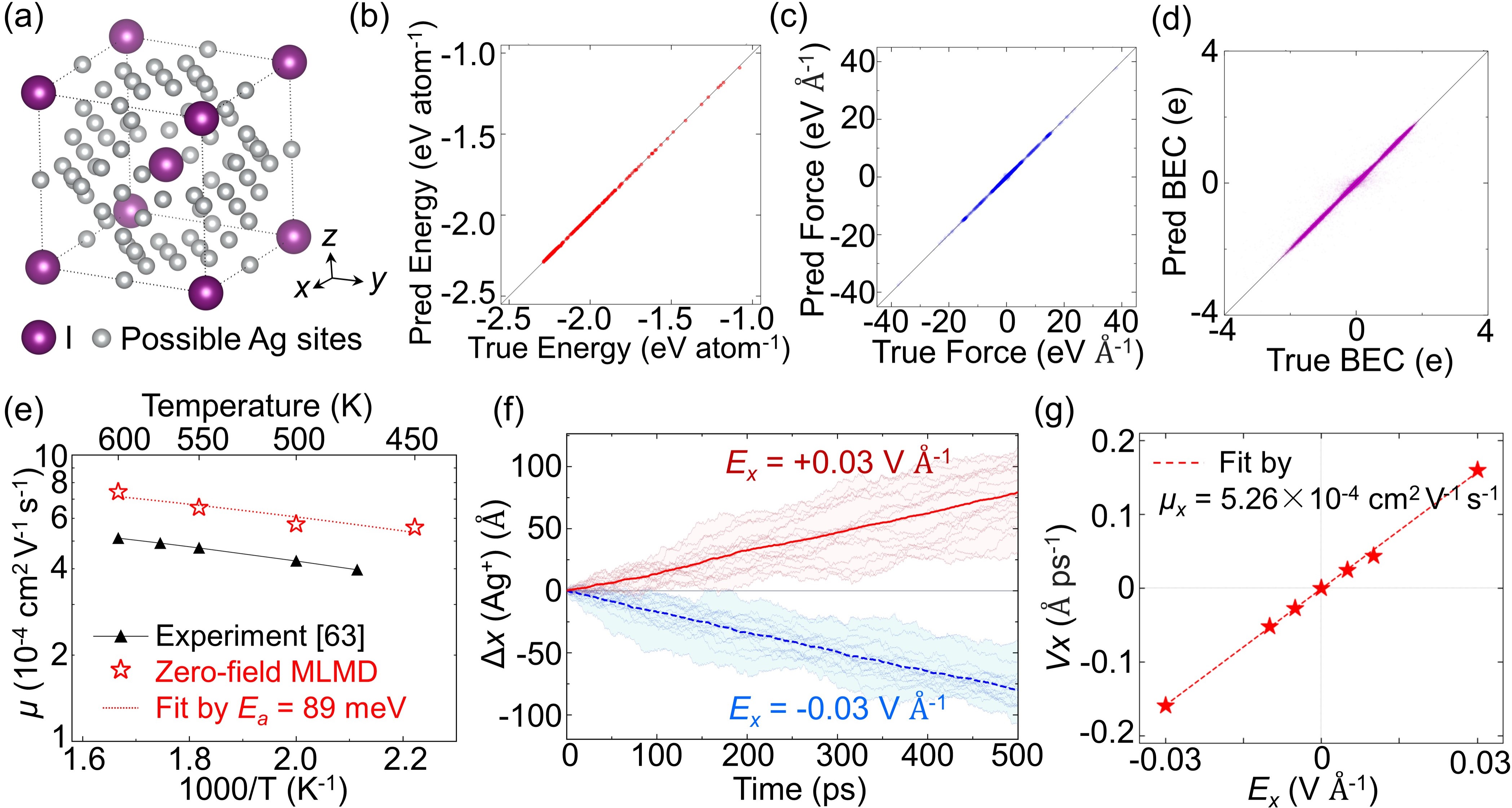}\\
	\caption{
		\textbf{EFR-GNN captures thermally activated Ag$^{+}$ transport and field-driven ionic drift in $\alpha$-AgI.}
		(a) Cubic structure of $\alpha$-AgI, showing the body-centered-cubic I framework and accessible Ag$^+$ sites.
		(b--d) EFR-GNN predictions versus DFT references for energy, atomic forces, and BEC components, respectively.
		The corresponding MAEs are 0.85 meV atom$^{-1}$, 18.9 meV \AA$^{-1}$, and 0.027 e, respectively.
		(e) Zero-field Ag$^+$ tracer mobility from two independent 1-ns MLMD trajectories in the canonical ensemble at each temperature, compared with experiment \cite{Kvist1970}.
		The Arrhenius fit gives an activation energy of 89 meV.
		(f) Ag$^+$ displacements projected along $x$ at 600~K under $E_x=\pm0.03$~V\,\AA$^{-1}$; thin curves denote the displacement traces of individual Ag$^+$ ions, the shaded regions span the corresponding minimum-to-maximum ranges, and the thick curves show the mean Ag$^+$ displacement.
		(g) Mean drift velocity versus electric field at 600~K, linearly fitted to give $\mu_x=5.26\times10^{-4}$~cm$^2$\,V$^{-1}$\,s$^{-1}$, in good agreement with the zero-field tracer mobility in (e).
	}
	\label{Fig6}
\end{figure*}

Next, we apply EFR-GNN to silver iodide (AgI), an archetypal type-I superionic conductor that undergoes a transition near 420~K from the low-temperature $\beta$/$\gamma$ phases to the cubic $\alpha$ phase \cite{Wood2006,Hajibabaei2025}. 
As illustrated in Fig.~\ref{Fig6}(a), a long-range-ordered iodine framework coexists with a dynamically disordered Ag$^+$ sublattice in $\alpha$-AgI, underlying its liquid-like Ag$^+$ diffusion and high ionic conductivity \cite{Hull2004}.
The fast Ag$^+$ transport has further enabled reversible ionic doping of two-dimensional semiconductors and programmable electronic devices \cite{Lee2020}.

A central challenge in modeling $\alpha$-AgI arises from the extensive configurational diversity of its dynamically disordered Ag$^+$ sublattice.
To achieve sufficiently broad and diverse sampling, DFT-MD trajectories were generated at 600~K from 500 independently initialized supercells.
A trajectory-level train--validation--test split was further adopted to ensure that all test trajectories originated from initial configurations not encountered during training.
EFR-GNN achieves low MAEs of 0.85 meV atom$^{-1}$ for energy, 18.9 meV \AA$^{-1}$ for atomic forces, and 0.027 $e$ for the BEC tensors, as presented in Fig.~\ref{Fig6}(b--d). 
Its energy and force accuracy is comparable to that of leading MLIPs developed for $\alpha$-AgI \cite{Vandermause2020,Hajibabaei2025}.
Detailed model parameters and evaluations are shown in Supplementary Note 8 \cite{SI}.

To evaluate thermally activated Ag$^+$ transport, we performed zero-field NVT MLMD simulations from 450 to 600 K. 
A 32-atom supercell was used.
At each temperature, two independent 1-ns trajectories were generated from distinct initial Ag$^+$ configurations.
Tracer diffusion coefficients were extracted from linear fits to the time-origin-averaged Ag$^+$ MSDs over lag times of 5--50 ps and converted to tracer mobilities through the Nernst--Einstein relation \cite{Carvalho2022}.
As displayed in Fig.~\ref{Fig6}(e), the predicted mobilities are on the order of $10^{-4}$~cm$^2$\,V$^{-1}$\,s$^{-1}$ and correctly reproduce the observed increase with temperature, although the absolute values remain systematically above experiment. 
The corresponding Arrhenius analysis yields an activation energy of 89 meV, consistent with the experimental value of approximately 95 meV \cite{Kvist1970}.
Detailed MLMD settings and analyses of trajectory stability, Ag$^+$ diffusion, and field-driven drift are shown in Supplementary Notes 9 and 10 \cite{SI}.

To examine field-driven ion transport, static electric fields of $E_x=\pm0.005$, $\pm0.01$, and $\pm0.03$~V\,\AA$^{-1}$ were applied along the $x$ direction at 600~K.
For each field strength, 500-ps NVT MLMD simulations were performed.
As shown in Fig.~\ref{Fig6}(f), opposite fields of $\pm0.03$ V \AA$^{-1}$ produce sustained Ag$^+$ drift in opposite directions, whereas the broad distribution of individual displacements reflects the thermal fluctuations superimposed on the collective motion. 
The drift velocities obtained from linear fits to the mean Ag$^+$ displacements vary approximately linearly with the electric field, as displayed in Fig.~\ref{Fig6}(g).
A linear fit yields a drift mobility of $\mu_x=5.26\times10^{-4}$ cm$^2$ V$^{-1}$ s$^{-1}$, consistent in magnitude with both the zero-field tracer mobility and the experimental tracer mobility at 600 K \cite{Kvist1970}.
The reversal of the drift direction with field polarity and its near-linear field dependence indicate sustained collective Ag$^+$ transport over the investigated field range.

\section*{Discussion}

We have developed EFR-GNN for long-time finite-temperature dynamics under static and time-dependent electric fields.
By combining a field-free interatomic potential with atom-resolved electronic-state descriptors and configuration-dependent BECs, the framework extends field-responsive MLMD to processes involving both collective atomic motion and atom-resolved charge degrees of freedom.
In hole-doped MgO, the model follows the microscopic dynamics of a hole polaron and reveals how an electric field breaks the directional balance of thermally activated hopping, rectifying stochastic intersite motion into sustained drift.
In GaAs, it resolves the excitation and subsequent decay of a coherent polar Γ-TO phonon and further shows deterministic control of its rotational handedness through the helicity of a circularly polarized THz pulse. 
In superionic $\alpha$-AgI, the framework captures the experimental temperature trend and activation energy of Ag$^+$ mobility and simulates collective long-range ionic drift under a static electric field. 
These applications span complementary classes of field-driven dynamics at finite temperature.

The present framework has a well-defined range of validity.
EFR-GNN propagates atomic motion on a field-free ground-state potential-energy surface, with field-induced forces introduced through configuration-dependent BECs. 
The results should therefore be interpreted within a configuration-dependent linear-response approximation, in which the electric field modifies atomic forces through the instantaneous BEC tensors without nonlinearly reshaping the underlying potential-energy surface. 
The predicted atomic charges and local magnetic moments are structure-dependent descriptors rather than independently propagated electronic degrees of freedom and are not explicitly conditioned on the applied field. 
Consequently, the framework captures electronic-state changes accompanying field-driven structural evolution, but not field-induced charge redistribution or carrier relocalization at fixed geometry. 
Purely electronic excitations and nonadiabatic dynamics remain beyond the present scope and would require an explicitly field-dependent electronic representation or time-dependent electronic propagation.

\section*{Methods}

\subsection*{Density functional theory calculations and construction of datasets }

All density-functional-theory (DFT) calculations were performed using the Vienna \textit{Ab initio} Simulation Package (VASP) \cite{Kresse1996}. 
The electron--ion interactions were described using the projector-augmented-wave method \cite{Bloechl1994}, and exchange-correlation effects were treated within the Perdew--Burke--Ernzerhof (PBE) generalized-gradient approximation \cite{Perdew1996}. 
The plane-wave cutoff energy was set to 600 eV. 
The electronic self-consistency threshold was $10^{-5}$ eV. Structural optimizations were continued until the residual force on each atom was below 10$^{-2}$~eV~\AA$^{-1}$.
Atom-resolved charges were obtained by Bader partitioning of the charge density \cite{Henkelman2006}, and the Born effective charge tensors were calculated using density-functional perturbation theory (DFPT) \cite{Baroni2001}.

For hole-doped MgO, spin-polarized PBE+$U$ calculations were performed using the rotationally invariant Dudarev formulation \cite{Dudarev1998}, with $U_{\mathrm{eff}}=10$~eV applied to the O $2p$ orbitals and no Hubbard correction applied to Mg \cite{Falletta2022,Birschitzky2025}. 
A hole was introduced by removing one electron from a 64-atom $2\times2\times2$ MgO supercell, together with a uniform neutralizing background charge. 
To generate a symmetry-broken initial configuration for the localized hole polaron, a selected O atom was temporarily replaced by S, the structure was relaxed, and S was subsequently replaced by O while retaining the induced local distortion. 
The nearest-neighbor O--O hopping pathway was initialized by linear interpolation between the relaxed endpoint structures and subsequently optimized using the nudged elastic band method implemented in VASP \cite{Henkelman2000}.
Three first-principles NVT trajectories were generated at 500 K using a 1-fs timestep, a Nos\'e thermostat, and $\Gamma$-point sampling, with the polaron initially localized on three different O sites. 
For each hopping event, identified by a change in the O site carrying the maximum local magnetic moment, 20 configurations on either side of the transition were retained. 
Together with configurations sampled along a nearest-neighbor O--O hopping pathway, this procedure yielded approximately 6,600 structures labeled with energies, forces, Bader charges, magnetic moments, and DFPT Born effective charge tensors. 
The dataset was divided into training, validation, and test subsets in a 70:15:15 ratio.

For GaAs, a 4-ps first-principles NVT trajectory was generated at 600 K using a 16-atom $2\times2\times2$ supercell, a 1 fs timestep, a Nos\'e thermostat, and a Γ-centered $3\times3\times3$ $k$-point mesh. 
All 4,000 molecular-dynamics configurations were used directly to construct the energy--force dataset. 
From the same trajectory, 800 uniformly spaced configurations were selected for DFPT calculations using the same $k$-point mesh to provide the BEC tensors. 
The datasets were divided into training, validation, and test subsets in a 70:15:15 ratio.

For $\alpha$-AgI, 500 independent 1-ps first-principles NVT trajectories were generated at 600 K using 32-atom supercells with distinct initial Ag$^{+}$ arrangements, a 2 fs timestep, and a Nos\'e thermostat. 
Ten uniformly spaced configurations were extracted from each trajectory, yielding approximately 5,000 structures. 
For each configuration, energies, forces, charge densities, and Born effective charge tensors were calculated using a Γ-centered $2\times2\times2$ $k$-point mesh. 
To avoid correlations between related configurations, the trajectories were divided into training, validation, and test sets in a 70:15:15 ratio before snapshot extraction, resulting in approximately 3500, 750, and 750 structures, respectively, with no trajectory shared across the three subsets.

\section*{Acknowledgements}
This work was supported by the start-up funding of the National Center for Nanoscience and Technology and by the National Natural Science Foundation of China (Grant No. 12574255).
Xinghua Shi is supported by National Key R\&D Program of China (2022YFA1203200), the National Natural Science Foundation of China 12125202.
Sheng Meng acknowledges support from the Chinese Academy of Sciences (YSBR-047) and the National Natural Science Foundation of China (12450401).

\section*{Author contributions statement}

Jia-Wen Li and Jin Zhang conceived the original ideas and supervised the work. 
Jia-Wen Li performed the first-principles calculations, code development and data analysis.
All authors participated in discussing and editing the manuscript.

\section*{Additional information}

\noindent\textbf{Data availability.}
The data supporting the findings of this study are available from the corresponding authors upon reasonable request.

\medskip

\noindent\textbf{Competing interests.}
The authors declare no competing interests.

\end{document}